\documentclass[pre, reprint,twocolumn,showpacs,preprintnumbers,amsmath,amssymb,a4paper]{revtex4-1}
\usepackage{graphicx}
\usepackage{epsfig}
\usepackage{natbib}
\usepackage{amsfonts}
\usepackage{wasysym}
\providecommand{\abs}[1]{\lvert#1\rvert}

\begin{document}

\title{Formation of localized structures in bistable systems through nonlocal
spatial coupling I: General framework}

\author{Pere Colet$^1$, Manuel A.~Mat\'{\i}as$^1$, Lendert Gelens$^{1,2}$, and Dami\`a Gomila$^{1}$}
\affiliation{
$^1$IFISC, Instituto de F\'{\i}sica Interdisciplinar y Sistemas 
Complejos (CSIC-UIB), Campus Universitat Illes Balears, E-07122 Palma de
Mallorca,  Spain \\
$^2$Applied Physics Research Group (APHY), Vrije Universiteit Brussel, Pleinlaan 2,
B-1050 Brussel, Belgium}
\date{\today}


\begin{abstract}
The present work studies the influence of nonlocal spatial coupling on the existence of localized structures in 1-dimensional extended systems. We consider systems described by a real field with a nonlocal coupling that has a linear dependence on the field. 
Leveraging spatial dynamics we provide a general framework to understand the effect of the nonlocality on the shape of the fronts connecting two stable states. In particular we show that nonlocal terms can induce spatial oscillations in the front tails, allowing for the creation of localized structures, emerging from pinning between two fronts. In parameter space the region where fronts are oscillatory is limited by three transitions: the modulational instability of the homogeneous state, the Belyakov-Devaney transition in which monotonic fronts acquire spatial oscillations with infinite wavelength, and a crossover in which monotonically decaying fronts develop oscillations with a finite wavelength. We show how these transitions are organized by codimension 2 and 3 points and illustrate how by changing the parameters of the nonlocal coupling it is possible to bring the system into the region where localized structures can be formed.
\end{abstract}

\maketitle

\section{Introduction}

Classical evolution equations describing the dynamics of a field in space and
time are Partial Differential Equations (PDEs), like the heat and diffusion
equations. The spatial interaction is expressed in terms of derivatives of the
relevant field at each point, a local quantity. Alternatively, some systems can
be described considering that the coupling is global (i.e.
all-to-all), and both local and global nonlinear evolution equations often
display complex behavior \cite{Cross_RevModPh_1993,Kuramoto_Springer_1984}.
More recently (see Ref. \cite{Murray_Springer_2002} for a survey), considerable effort 
has been devoted to the study of evolution equations in which the spatial 
interaction is nonlocal, intermediate between local and global, being the 
spatial interaction written in the form of an integral over an spatial domain, 
leading to an integro-differential equation. These spatially nonlocal effects 
are known to be relevant in a number of fields, ranging from chemical reactions  
\cite{Kuramoto_PRL_1998, *Shima_PRE_2004, *Nicola_nonlocal02}, to several
problems in Biology and Ecology \cite{Murray_Springer_2002}, including
Neuroscience \cite{Ermentrout_RPP,*Coombes05} (with examples like neural
networks underlying mollusk patterns \cite{Ermentrout_Shells,*Boettiger_PNAS,*Gong_PNAS} and
hallucination patterns \cite{ErmentroutCowan79}) and population dynamics  
\cite{Fuentes_PRL_2003,*2004PhRvE_Hernandez,*Clerc_PRE_2005,Clerc_PRE_2010}. Some mechanisms
through which an effective nonlocal interaction may emerge are a physical/chemical interaction 
that couples points far apart in space, e.g., a long-range interaction \cite{PedriSantos_PRL_2005}, 
or from the adiabatic elimination of a slow variable \cite{Kuramoto_PTP_1995,Kuramoto_PRE_2003}. Novel
phenomena emerging genuinely from nonlocality, such as power-law correlations
\cite{Kuramoto_PTP_1995,Kuramoto_PRL_1996},  multiaffine turbulence
\cite{Kuramoto_PRL_1998}, and chimera states \cite{Kuramoto_chimera_2002,*Abrams_PRL_2004,
*Sethia_PRL_2008} have been reported. Moreover, recent works have reported the
effects of nonlocality on the dynamics of fronts, patterns and localized 
structures (LSs), for instance the tilting of snaking bifurcation lines 
\cite{Firth_PRL_2007} and changes in the size of LSs \cite{Gelens_PRA_2007,*Gelens_PRA_2008}, the
effects of two-point nonlocality on convective instabilities 
\cite{Zambrini_PRL_2005}, the nonlocal stabilization of vortex beams
\cite{Skupin-2007},  or changes in the interaction between solitons \cite{Neshev},
and in the velocity of propagating fronts \cite{Maruvka2007}. 

A situation particularly interesting to study the effects of nonlocal interactions arises in systems far from equilibrium with fronts connecting two coexisting stable homogeneous steady states (HSS). If the system is variational, i.e, it derives from a potential, the front will move such that the most stable state will invade the least stable (metastable) one, and the (uniquely defined) front velocity can be related to the difference in potential between the two states \cite{Pomeau86}. A bit more subtle is the case that the two HSS are equivalent. In principle, none of them will prevail and a front will not move, although if 
several fronts coexist in the system, short-range tail interactions come into play. These short-range forces may also be relevant in the case that the difference
in relative stability is small.
These interaction forces decay exponentially, being attractive for monotonic fronts
\cite{Coullet87} and, thus, two fronts (a kink and an anti-kink) tend to annihilate 
mutually \footnote{At least in 1-D, because in 2-D curvature effects play an important 
role.}. If the fronts exhibit oscillatory tails, a kink-antikink pair may lock 
\cite{Pomeau86, Knobloch08}, potentially leading to a chaotic sequence (spatial chaos) \cite{Coullet87}.
As a result of the locking of a kink-antikink pair, a LS is obtained. 
These LSs, emerging from the interaction of two equivalent homogeneous states, are different 
from those arising through the interaction of a homogeneous state and a pattern appearing 
subcritically \cite{Fauve,*TlidiTKrev07,*WoodsChampneys,*Coullet00}.

In Ref.\ \cite{GelensPRL2010} we considered the role of a nonlocal spatial
coupling in the interaction of fronts connecting two equivalent states in $1$-D
spatially extended systems.
In the local case the interaction between two fronts (a kink and an anti-kink) decays 
exponentially with distance \cite{Coullet87}. In the nonlocal case, at least if the kernel decays 
exponentially or faster \footnote{Nonlocal interaction terms decaying slower than exponential has been considered for example in Ref.~\cite{Escaff_EPJD_2010,*Clerc_PRL_2013}}, front interactions still decay exponentially.
In \cite{GelensPRL2010} it was shown that a nonlocal interaction enhances the interaction, extending 
substantially the range of interaction.
Moreover, in \cite{GelensPRL2010} another effect was observed for the case of
repulsive spatially nonlocal interactions of systems exhibiting monotonic fronts,
namely the appearance of spatial tails, leading to stable LSs. Repulsive (inhibitory) 
interactions are common, for instance, in neural field theories 
\cite{Ermentrout_RPP,*Coombes05} and genetic networks \cite{Alonbook}.
In particular, this effect was found for the real Ginzburg-Landau equation,
which does not exhibit tails with a local interaction, subject to a Gaussian 
spatially nonlocal kernel. This result is generic and can be qualitatively understood from
the interplay between nonlocality, which couples both sides of the front, and repulsiveness 
which induces a small depression at the lower side and a small hill at the upper part. 

The goal of the present manuscript is to provide a general framework to understand the effect of nonlocal spatial coupling on the shape of the fronts connecting stable steady states. This allows determining the parameter regions in which fronts have oscillatory tails and therefore LSs can exist. 
Leveraging spatial dynamics we obtain general results for 1-D extended systems described by a real field with nonlocal coupling terms that are linear in the field. 
In a second manuscript \cite{PartII}, which we will refer hereafter as Part II, we apply the theoretical analysis developed here to the real Ginzburg-Landau equation with different spatial kernels, including the Gaussian which is positive definite kernel and decays in space faster than exponentially, the mod-exponential kernel, also positive definite but with exponential decay, and a Mexican-hat shaped kernel which is non positive definite and thus has attractive and repulsive regions.

In terms of spatial dynamics the shape of the front is given by the leading complex eigenvalues which are the zeros of the spatial dispersion relation. It turns out that in order to properly describe the effect of nonlocal kernels with attractive and repulsive regions, such as the Mexican-hat, one needs to consider at least six spatial eigenvalues, thus the minimal dispersion relation is a six order one. Therefore in this manuscript we address the different scenarios that one can encounter with six spatial eigenvalues.

In parameter space the region of existence of oscillatory tails is limited, on one side, by the onset of spatial oscillations on a monotonic front and, on the other side, by the oscillations becoming undamped, namely by the modulational instability (MI) of the HSS. The onset of spatial oscillations can, in fact, take place in two ways, a Belyakov-Devaney (BD) transition \cite{Devaney76,Homburg10} in which oscillations appear initially at infinite wavelength, and a crossover in which finite wavelength oscillations develop. 

These three transitions are codimension 1 (codim-1) manifolds, that is, they have one dimension less than the space of parameters. They are organized in such a way that the region where LSs can exist unfolds from two codim-2 points. 
One is a local bifurcation in which the dispersion relation has a quadruple zero and from which MI and BD manifolds unfold in opposite directions and then bend in a parabolic way \cite{Champneys98,Haragus11}. In the parameter region between these two manifolds fronts have oscillatory tails. 
The other is a nonlocal transition in which the HSS becomes simultaneously unstable to homogeneous and to finite wavelength perturbations. MI and crossover manifolds unfold from this transition, one secant to the other, and fronts have oscillatory profiles in the parameter region between them. 

Furthermore, there is another codim-2 bifurcation that plays a relevant role in the overall phase space organization. It is a cusp where two BD manifolds originate (or end) tangentially one to the other and which also unfolds a crossover manifold. In particular in the first of the parameter regions described above, the BD manifold unfolding from the quadruple zero ends at that cusp. After that, the transition from monotonic to oscillatory fronts that limits the parameter region where LSs can exist is given by the crossover manifold. 

All these codim-2 bifurcations unfold from a codim-3 local bifurcation point characterized by being a sextuple zero of the dispersion relation, which, to the best of our understanding, has not been characterized previously.

When the dynamics includes the effect of nonlocal interaction terms often the dispersion relation becomes a transcendental function with an infinite number of zeros. By playing with the parameters of the nonlocal interaction it is possible to bring the system into the parameter region where fronts have spatially oscillatory tails allowing for the existence of LSs. Nonlocal interaction terms can also induce the opposite effect, namely, to preclude the formation of LSs in systems in which they are present.

The manuscript is organized as follows. In Section \ref{Sect::System}, we introduce 
the generic spatially extended systems under study. In Section \ref{Sect::KernelTransformations} we consider 
transformations of the nonlocal interaction term which allow for systematic approximations.
In Section \ref{Sect::linstab} we analyze the effect of the nonlocal kernel in the temporal stability of HSS. In Section \ref{Sect::SpatDyn} we describe the HSS in terms of the spatial dynamics and in Section \ref{Sect::LS_SpatDyn} we show how the spatial eigenvalues determine the shape of the front tails and thus the existence of LSs. Section \ref{Sect::transitions_LS} describes the MI, BD and crossover transitions. The quadruple-zero and the cusp are discussed in Sections \ref{Sect::QZ} and \ref{Sect::cusp}. Section \ref{Sect::SZ} is devoted to the sextuple zero codim-3 bifurcation.
Finally in Section \ref{Sect::conclusions} we illustrate how by changing the nonlocal interaction parameters it is possible to bring the system into the parameter regions in which fronts have an oscillatory profile.

\section{System}\label{Sect::System}

In the present section we describe the 1-D spatially extended systems and the nonlocal interaction terms considered in this manuscript. We start with a generic system with local coupling of the form 
\begin{equation}
\partial_t A  = G(A,\partial_{xx}A, \partial_{xxxx}A, ...)
\label{eq:generic}
\end{equation}
where $A \equiv A(x,t)$ is a real field and $G$ is a nonlinear function of the field and its even-order spatial derivatives. The system is translationally invariant and, thus, $G$ is symmetric under the parity transformation $x\leftrightarrow -x$.
We will consider that the system has several HSSs, $A_s$, for which $G(A_s,0,0,...)=0$.

Now let us consider an extension of the systems considered above to include a spatially nonlocal term $F(x,\sigma)$
\begin{equation}
\partial_t A  = G(A,\partial_{xx}A, \partial_{xxxx}A, ...) -sM_0A + sF(x,\sigma),
\label{eq:generic_nonlocal}
\end{equation}
where $s$ is a parameter regulating the strength of the nonlocal term.
Thus Eq.~(\ref{eq:generic_nonlocal}) has two spatial interaction terms: a
local (diffusive or higher order) and a nonlocal spatial coupling. In this work we consider that the spatially nonlocal term is linear in the field $A$ and is defined through the convolution of a spatially nonlocal influence function (or kernel), $K_{\sigma}(x)$, with the local field $A(x)$:
\begin{equation}
F(x,\sigma) =  \int_{-\infty}^{\infty} \! K_{\sigma}(x-x')\,A(x')dx'\, ,
\label{eq:nonlocaldef}
\end{equation}
where the parameter $\sigma$ controls the
spatial extension (width) of the coupling. We will consider that the kernel, and thus the nonlocal term, preserves the symmetry under the parity transformation $x\leftrightarrow -x$, namely $K_{\sigma}(x)=K_{\sigma}(-x)$.
The nonlocal
term $F(x,\sigma)$ has a local contribution that is compensated for by the term
$-s M_0 A$, where $M_0 = \int_{-\infty}^{\infty} K_{\sigma}(x) dx$.
The HSSs of the nonlocal system (\ref{eq:generic_nonlocal}) are the same as the original system with local coupling 
(\ref{eq:generic}). This is because for any HSS $A(x)=A_s$, $F(x,\sigma) = M_0 A_s$ which is canceled out by the $-s M_0 A$ term.
If the kernel $K_{\sigma}(x)$ does not crosses zero, without loss of generality we take $K_{\sigma}(x)$ to be positive definite. Then for $s>0$ the interaction is attractive for all distances $x$. Conversely the interaction is repulsive for $s<0$. If $K_{\sigma}(x)$ crosses zero then the kernel has attractive and repulsive regions, as is the case for the Mexican-hat kernel to be considered in Part II.

A relevant consequence of the symmetry of the kernel is that the nonlocal term
(\ref{eq:nonlocaldef}) can be derived from a nonequilibrium potential. The
nonequilibrium potential ${\cal F}$ of the nonlocal part is:
\begin{equation}
{\cal F}[A(x),A(x')]=\int_{-\infty}^{\infty}\!
\int_{-\infty}^{\infty} A(x) K_{\sigma}(x-x') A(x') dx\,dx'
\label{eq:nonlocalpot}\ .
\end{equation}
Therefore, if the original system was variational, the introduction of the nonlocality in the form considered here preserves the variational character of the system.  The variational character has important implications \cite{Pomeau86}, like the fact that the many possible solutions of the problem are local minima (metastable states)
of a functional and cannot exhibit neither temporal oscillations nor the so called Nonequilibrium Ising-Bloch 
transition \cite{Coullet90}, by which chiral fronts may spontaneously start to move, being both manifestations
of nonvariationality.

For later convenience we define the Fourier transform of the nonlocal kernel,
\begin{equation}
\hat{K}_{\sigma}(k) =  \int_{-\infty}^{\infty} \! K_{\sigma}(x)\,e^{-ikx} dx \, . 
\label{eq:Kernel_Fourier}
\end{equation}
In what follows we will use the hat symbol $\hat{\ }$ to label functions in Fourier space.
Since $K_{\sigma}(x)$ is real and $K_{\sigma}(x)=K_{\sigma}(-x)$,  $\hat{K}_{\sigma}(k)$ is also real and symmetric with respect to the transformation $k\leftrightarrow -k$, namely $\hat{K}_{\sigma}(k)=\hat{K}_{\sigma}(-k)$ \cite{Abramowitz}. Therefore $\hat{K}_{\sigma}(k)$ depends on $k$ only through $k^2$ and we can write,
\begin{equation}
 \hat{K}_{\sigma}(k) = \tilde{\hat{K}}_{\sigma}(u) \, ,
 \label{Eq::K_tilde}
\end{equation}
where  $u=k^2$.

Using the convolution theorem for the Fourier transform, in Fourier space the interaction term can be written as,
\begin{equation}
 \hat{F}(k,\sigma) =  \int_{-\infty}^{\infty} \! F(x,\sigma)\,e^{-ikx} dx=     
\hat{K}_{\sigma}(k)\,\hat{A}(k)\ . 
\label{eq:nonlocaltran}
\end{equation}

\section{Kernel transformations and expansions}
\label{Sect::KernelTransformations}

The Fourier formalism is very useful as it allows to perform exact transformations of the nonlocal interaction term as well as to develop systematic approximations. In many instances the nonlocal kernel $\hat{K}_{\sigma} (k)$ is a transcendental function. In order to obtain approximate equations for the dynamics one may consider a Taylor expansion of the nonlocal kernel $\hat{K}_{\sigma} (k)$ around $k=0$. If $\hat{K}_{\sigma} (k)$ has singularities in the complex plane, the location of the singularity closest to the origin determines the radius of convergence of the expansion. Then one can resource to a Laurent expansion around the singularities which allows to derive a differential equation for the nonlocal interaction term involving only spatial derivatives. In the next two subsections we consider the Taylor and the Laurent expansions respectively.

\subsection{Moment Expansion}
\label{Sect::MomentExpansion}

Assuming $\hat{K}_{\sigma} (k)$ has no singularities at finite distance and considering that because of the symmetries the expansion of $\hat{K}_{\sigma} (k)$ has only even powers of $k$ one can write,
\begin{eqnarray}
\hat{K}_{\sigma} (k) &=& \sum_{j=0}^{\infty} (-1)^j \frac{M_{2j}}{(2j)!}  k^{2j},
\end{eqnarray}
where,
\begin{equation}
 M_{2j}= (-1)^j \left. \frac {d^{2j} \hat{K}_{\sigma}(k)}{dk^{2j}}\right|_{k=0}=\int_{-\infty}^{\infty} x^j K_{\sigma}(x) dx \, ,
\end{equation}
are the moments of the nonlocal kernel. Using this kernel expansion the nonlocal interaction in real space can be written as,
\begin{align}
F(x,\sigma) &= \frac{1}{2 \pi}  \int_{-\infty}^{\infty} e^{ikx} \sum_{j=0}^{\infty} (-1)^j \frac{M_{2j}}{(2j)!} k^{2j} \hat A (k) dk \nonumber \\
& = \sum_{j=0}^{\infty} \frac{M_{2j}}{(2j)!}  \frac{\partial^{2j} A}{\partial x^{2j}} \ .
\label{eq::taylorexp}
\end{align}
In real space, the result of the transformation is that one expands a spatially nonlocal term as a series of spatial derivatives of $A$ of even order \cite{Murray_Springer_2002}, that formally yields a sum, in principle an infinite one, of local contributions. A truncation of the series to order $2j$ is only mathematically justified if it converges fast enough, in other words if $\abs{M_{2(j+1)}} << \abs{M_{2j}}$. 

Nonlocal kernels typically decay to zero for large $k$. Therefore the effect of long wavenumber perturbations ($k \rightarrow \infty$) in (\ref{eq:generic_nonlocal}) is the same as in the system with only local coupling (\ref{eq:generic}). In models describing physical, chemical or biological systems these perturbations are damped. However, performing a moment expansion can introduce spurious instabilities if, after truncation, long wavenumber perturbations are amplified by the higher order term. To avoid these spurious instabilities it is necessary that the coefficient of the higher order term satisfies
\begin{equation}
 s (-1)^j M_{2j} <0.
\end{equation}
For positive definite kernels which have all the moments positive, this condition implies that for $s>0$ the expansion can only be truncated at order $2j$ with $j$ odd, namely at order $2(2m+1)$ with $m$ integer. Conversely for $s<0$ the expansion can only be truncated at order $4m$ with $m$ integer.

For nonlocal kernels that can be written in the form 
\begin{equation}
 K_{\sigma}(x)= (1/\sigma) K(x/\sigma) \, ,
 \label{Eq::scaled_kernel}
\end{equation}
the moments are given by,
\begin{equation}
 M_j= \int_{-\infty}^{\infty} x^j \frac{K(x/\sigma)}{\sigma}  dx = \sigma^j \int_{-\infty}^{\infty} y^j K(y) dy 
 \equiv \sigma^j {\cal M}_j \, .
\end{equation}
Thus $M_0$ is independent of the width $\sigma$ and the moment of order $j$ scales with $\sigma$ to the power $j$.

\subsection{Kernels with singularities. Laurent expansion}
\label{Sect::LaurentExpansion}

If $\hat{K}_{\sigma} (k)$ has singularities in the complex plane, the expansion in moments is of limited use.
In this case it is possible to use an alternative approach as follows. For the sake of clarity we first consider a kernel that has a simple singularity located in the complex plane at $k^2=u=\xi$. Then it is possible to write a Laurent expansion \cite{Phillips} of the form,
\begin{equation}
\tilde{\hat{K}}(u)=\sum_{j=1}^n  \frac{b_j} {(u-\xi)^{j}} + \sum_{j=0}^{m} a_j (u-\xi)^j \, ,
\end{equation}
where the coefficients $b_j$ and $a_j$ can be calculated from a Cauchy's line integral \cite{Phillips}. The first sum is the principal part of the function 
$\tilde{\hat{{K}}}(u)$. If the singularity of the function at $u=\xi$ is not an essential one, then the principal part has a finite number of terms, namely the singularity is a pole of order $n$.
The second sum has the form of a Taylor expansion which has a finite number of terms if $\tilde{\hat{K}}(u)$
does not have an essential singularity at infinity. 

The nonlocal term can be written as
\begin{equation}
  \hat{F}(k,\sigma) = \left[ \sum_{j=1}^n \frac{b_j} {(k^2-\xi)^{j}} + \sum_{j=0}^{m} a_j (k^2-\xi)^j \right] A(k) \, . 
\end{equation}
Multiplying in both sides by $(k^2-\xi)^n$ and going to real space one gets a higher order spatial differential equation for the nonlocal term
\begin{align}
&(-\partial_{xx}-\xi)^n F(x,\sigma) = \sum_{j=1}^n b_j (-\partial_{xx}-\xi)^{n-j} A(x) \nonumber \\
& \quad \quad + \sum_{j=0}^{m} a_j (-\partial_{xx}-\xi)^{n+j}  A(x) \, .
\label{Eq::LaurentExp_DifEq}
\end{align}
Combining this with (\ref{eq:generic_nonlocal}) results in a transformed equation for the dynamics. The principal part leads to a differential equation for the nonlocal interaction term involving only spatial derivatives while the Taylor part, as before, leads to a series of spatial derivatives of $A$ of even order. If the Taylor part has a finite number of terms this procedure leads to an exact transformation. Otherwise, approximate equations for the dynamics can be obtained by truncating the series with the caveats discussed in the previous subsection.

The approach discussed here is very general and can be extended to any kernel with several singular points provided the singularities are not essential. In this case the kernel can be written as \cite{Phillips}
\begin{equation} 
\tilde{\hat{K}}(u)=C+\sum_{l=1}^L \sum_{j=1}^{n(l)} \frac{b_{lj}}{(u-\xi_l)^{j}} + \sum_{j=1}^{m} B_j u^j ,
\end{equation}
where $C$ is a constant, $L$ is the number of poles, $n(l)$ the order of pole $\xi_l$ and the coefficients $B_j$ correspond to the principal part of the kernel at the point of infinity. This series has a finite number of terms and proceeding as before one gets an equation for the nonlocal interaction which generalizes (\ref{Eq::LaurentExp_DifEq}) and can be combined with (\ref{eq:generic_nonlocal}) to get an exact transformation of the dynamics. On the other hand, as before, kernels with an essential singularity at infinity have an infinite number of terms in the Taylor part which can eventually be truncated to get an approximation for the dynamics.

\section{Linear stability analysis of a Homogeneous Steady State (temporal dynamics)} 
\label{Sect::linstab}

The linear stability of a HSS is analyzed by considering the effect of finite wavelength
perturbations, $A = A_s + \epsilon \exp{(\Gamma t + i k x)}$. Linearizing for small $\epsilon$ one obtains for Eq. (\ref{eq:generic_nonlocal}) the dispersion relation:
\begin{equation}
\Gamma (k) = \Gamma_G (k) + s (\hat{K}_{\sigma}(k)-M_0) .
\label{Eq::dispersion_relation}
\end{equation}
where $\Gamma_G(k)$ is the dispersion relation obtained from linearization of $G$ around the HSS. Since $G$ is a real function symmetric under the transformation 
$x\leftrightarrow -x$, $\Gamma_G(k)=\Gamma_G(-k)$ \cite{Abramowitz}. So, the overall dispersion relation fulfills $\Gamma (k)=\Gamma (-k)$, and thus depends on $k$ only through $k^2$. Therefore one can write it in terms of $u=k^2$,
\begin{equation}
 \Gamma (k) = \tilde \Gamma (u) =  \tilde \Gamma_G (u) + s (\tilde{\hat{K}}_{\sigma}(u)-M_0) \, . 
 \label{Eq:Gamma_tilde}
\end{equation}

The HSS undergoes an instability if the maximum of $\Gamma (k)$ becomes positive when varying a system parameter. Therefore to have an instability at $k_c$ it is necessary that $\Gamma (k)$ has an extremum crossing 0, namely,
\begin{equation}
 \Gamma'(k_c) \equiv \left . \frac{d \Gamma(k)}{dk} \right|_{k_c} =0 \, , \quad \Gamma (k_c)=0 \, .
  \label{Eq::DZ_conditions_k}
\end{equation}
These conditions are precisely the conditions for $\Gamma (k)$ to exhibit a double zero (DZ) at $k_c$, namely a zero with multiplicity two. If $k_c\neq0$, then owing to the symmetry of $\Gamma (k)$, $-k_c$ is also a DZ, in other words, DZs at finite $k_c$ come in pairs. In order for a DZ to indeed signal the onset of an instability it is
necessary that $k_c$, besides being a local extremum, is the global maximum of $\Gamma (k)$.  

Due to the symmetry, $\Gamma'(0)=0$, the dispersion relation has always a local extremum at the origin. 
If the extremum at the origin corresponds to the absolute maximum, changing the constant term in $\Gamma(k)$ the HSS will eventually encounter a homogeneous instability.

Changing parameters different from the constant term may lead to instabilities at a finite $k_c$. In this case the HSS undergoes a modulational instability (MI), also referred to as Generalized Turing bifurcation \cite{Cross_RevModPh_1993}. 

In what follows it would be useful to consider the instabilities in terms of $\tilde\Gamma(u)$. Since, 
\begin{equation}
\frac{d \Gamma (k)}{dk}  =  \frac{du}{dk} \frac{d \tilde \Gamma (u)}{du} = 2k \tilde \Gamma'(u) \, , \nonumber
\end{equation}
a pair of DZs of $\Gamma(k)$ taking place at a $\pm k_c\neq0$ correspond to a DZ of $\tilde \Gamma(u)$ at $u_c=k_c^2$, 
\begin{equation}
\tilde \Gamma'(u_c)=0 \, ,  \quad \tilde \Gamma (u_c)=0 \, .
\label{Eq::DZ_conditions_u}
\end{equation}
On the other hand a DZ of $\Gamma(k)$ at $k_c=0$ does not correspond to a DZ in $\tilde \Gamma (u)$, rather it corresponds to a simple zero located at $u=0$, namely $\tilde \Gamma (0)=0$. 

In general the nonlocal kernel $\tilde{\hat{K}}_{\sigma}(u)$ can reshape the dispersion relation and can induce or damp MIs. Since in the dispersion relation (\ref{Eq:Gamma_tilde}) the term $s(\tilde{\hat{K}}_{\sigma}(u)-M_0)$ vanishes at $u=0$, the nonlocal coupling has no effect on homogeneous perturbations, only in finite wavelength ones. Note that while DZs of $\Gamma(k)$ can arise through either homogeneous or finite wavelength perturbations, DZs of $\tilde \Gamma (u)$ arise only due to finite wavelength perturbations, the ones the nonlocal coupling is acting on.

\section{Spatial dynamics} 
\label{Sect::SpatDyn}

When one intends to describe spatio-temporal structures
that are stationary in time, then, as the time derivative of the field is zero one
obtains a set of equations for the spatial evolution ({\it Spatial
Dynamics}), which form a special kind of dynamical system in which space
plays the role usually played by time.
For our generic system (\ref{eq:generic_nonlocal}), defining a set of intermediate variables $V_i$ for $i=1,...,2n-1$ where $n$ is such that the higher order spatial derivative in $G$ is of order $2n$, then one obtains the following 2n-dimensional {\it spatial\/} dynamical system 
\begin{eqnarray}
A' &=&V_1 \nonumber\\
V_{i}'&=& V_{i+1}\, , \; \; i=1,...,2(n-1)\nonumber\\
0&=&G(A,V_2,V_4,....,V_{2(n-1)},V_{2n-1}') \nonumber \\
&& \; \; - sM_0A + sF(x,\sigma) \, ,
\label{eq:genericSpatDyn} 
\end{eqnarray}
where the prime symbol stands for derivatives with respect to the spatial variable $x$. 
The last equation is an implicit equation for $V_{2n-1}'$. Typically the higher order derivative in $G$ appears in an additive way, then the last equation can be readily written in an explicit form. The spatial dynamical system (\ref{eq:genericSpatDyn}) has the distinctive property of reversibility whose consequences have been studied thoroughly \cite{Champneys98,Devaney76}. More precisely in this work we will be dealing with even-dimensional reversible systems \cite{Champneys98}. The concept of reversibility can be extended to odd-order systems, see for example Ref.~\cite{Haragus11}.

The fixed points of (\ref{eq:genericSpatDyn}) correspond to solutions in which the field does not have any dependence on $x$, $A(x)=A_s$, namely to HSSs.
The linearized stability equation for the spatial dynamics can
be obtained by considering a spatial perturbation of the form $A(x) = A_s + \epsilon \exp(\lambda x)$, where, in general, $\lambda$ is complex. Since the linearization is around the same state as in the temporal stability analysis and since the perturbations are the same as the ones considered there replacing $ik$ 
by a complex $\lambda$, the spatial eigenvalues $\lambda_0$ satisfy
\begin{equation}
 \Gamma(-i\lambda_0)=0,
\end{equation}
where $\Gamma(-i\lambda)$ is the dispersion relation (\ref{Eq::dispersion_relation}) but with a complex argument.
For the sake of simplicity in the notation we define,
\begin{equation}
 \Gamma_s(\lambda) \equiv \Gamma(-i\lambda) \, .
\end{equation}

\section{Localized Structures in the context of spatial dynamics}
\label{Sect::LS_SpatDyn}

Spatial eigenvalues are given by the zeros of the dispersion relation $\Gamma_s(\lambda)$, which depends on $\lambda$ only through $\lambda^2=-u$, thus they can be obtained from $\tilde \Gamma(u_0)=0$. As a consequence, spatial eigenvalues come in pairs, each spatial eigenvalue $\lambda_0$ being accompanied by its counterpart $-\lambda_0$. To be more precise, if $u_0$ is a real zero of $\tilde \Gamma(u)$, then there is a doublet of spatial eigenvalues $\lambda_0=\pm\sqrt{-u_0}$ with $\lambda_0$ real for $u_0<0$ or purely imaginary for $u_0>0$. If $u_0$ is a complex zero of $\tilde \Gamma (u)$ then $u_0^*$ is also a zero and therefore complex spatial eigenvalues must come in quartets $\lambda_0=\pm q_0 \pm ik_0$. As a consequence 
all fixed points have both attracting and repelling directions. Thus, fronts connecting a fixed point with itself (homoclinic orbits) are generic for all HSSs. These homoclinic orbits correspond to LSs.
While in generic (i.e., not reversible) dynamical systems homoclinic orbits can possibly be found by varying one parameter  (i.e., they are of codimension-$1$), in even-dimensional reversible systems they are of codimension-$0$ (i.e, they are persistent) \cite{Devaney76,Champneys98}, allowing for the pervasive presence of LSs in extended systems \cite{Akhmediev_book05}.

Spatial dynamics allows for the description of the stationary profile of the field, in particular of profiles arising from the interaction of fronts connecting different HSSs. Two such fronts can lock, leading to a LS, if they exhibit oscillatory tails \cite{Pomeau86,Knobloch08,Burke06}, what in the perspective of spatial dynamics means that the linearization of (\ref{eq:genericSpatDyn}) around the HSSs must lead to a complex quartet of (spatial) eigenvalues. 
The existence of oscillatory tails in the front profile can be elucidated from the spatial dynamics close to the HSS.
For LSs to form the tail oscillation amplitude should be large enough to overcome the attracting dynamics of the fronts which depends on the global front profile. Locking is easier for fronts connecting two equivalent HSSs since the interaction between the fronts is weak. Locking in fronts connecting two non equivalent HSSs is also possible but the oscillation amplitude has to be large enough to overcome the difference in stability \cite{Knobloch08,Burke06}. 
Once fronts lock, a LS can be viewed as a homoclinic orbit biasymptotic to a HSS passing close to the other HSS. In the case of fronts between equivalent states, both HSS are related by a system symmetry. 

In general the nonlocal interaction changes the number of spatial eigenvalues as well as their values. While there can be an arbitrary large number of eigenvalues, if the eigenvalues are well separated, the leading eigenvalues, i.e. those with the smallest $\abs{{\rm Re}(\lambda_0)}$,
determine the qualitative behavior, as they determine the asymptotic approach
to the fixed points. This leads to three different cases,
\begin{itemize}
\item[A] The leading eigenvalues consist of a purely real doublet $\lambda_0=\pm q_0$. A front starting (or ending) in the HSS has 
monotonic tails. 
\item[B] The leading eigenvalues consist of a quartet of complex eigenvalues $\lambda_0=\pm q_0 \pm i k_0$. A front starting (or ending) in the HSS has oscillatory tails, where the spatial oscillation wavenumber is determined by $k_0$. 
\item[C] The leading eigenvalues consist of a purely imaginary doublet $\lambda_0=\pm i k_0$. This means $\Gamma(k_0)=0$ for a real $k_0$. As a consequence $\Gamma(k)$ must be positive either for $k>k_0$ or for $k<k_0$. Since there are values for $k$ for which the dispersion relation is positive, the HSS is temporally unstable. 
\end{itemize}
Now considering two fronts connecting two equivalent states placed back to back, in case A the fronts will move decreasing the distance between them. Such behavior is also called \textit{coarsening} behavior \cite{CahnHill}. In case B fronts can lock at their tails and form LSs \cite{Pomeau86}.

\section{Transitions leading to Localized Structures}
\label{Sect::transitions_LS} 

There are two transitions that bring the system to case B starting from A and one starting from C. They can be understood by considering the location of the spatial eigenvalues in the $({\rm Re}(\lambda),{\rm Im}(\lambda))$ plane.

The typical transition that leads to case B starting from A is the collision of two doublets on the real axis resulting in a complex quartet emerging off-axis. This is the Belyakov-Devaney (BD) transition \cite{Devaney76,Homburg10,Champneys98}, also known as reversible $0^{2+}$ \cite{Haragus11}. At the BD spatially monotonic fronts become oscillatory. The BD is not a bifurcation since it does not involve any eigenvalue crossing the imaginary axis. In terms of $\tilde \Gamma (u)$ it corresponds to the collision of two zeros on the real negative semi-axis for $u$, namely to a DZ of $\tilde \Gamma (u)$ taking place for $u_c$ real and negative.

The entrance in case A starting from C corresponds to two doublets of imaginary spatial eigenvalues colliding and leading to a complex quartet emerging just off the imaginary axis. This is a Hamiltonian-Hopf (HH) bifurcation \cite{Champneys98}, also known as reversible $0^{2+}(i\omega)$ \cite{Homburg10,Haragus11}. Looking at $\tilde \Gamma (u)$, this bifurcation is signaled by the collision of two zeros on the real positive semi-axis, namely a DZ at $u_c$ real and positive. This corresponds to two DZ, thus to two extrema, in $\Gamma(k)$ at finite $k_c=\pm \sqrt{u_c}$. If $k_c$ turns out to be the global maximum of $\Gamma(k)$, then the HH corresponds to a MI. 

Curiously enough, there is an additional transition to go from the case A to B which is not associated to a collision. It involves a real doublet $\lambda_1=\pm q_1$ leading the spatial dynamics and a complex quartet $\lambda_2=\pm q_2 \pm i k_2$ with $q_2>q_1$. If the quartet moves towards the imaginary axis when changing a parameter, $q_2$ will get closer to $q_1$ and the eigenvalues will no longer be well separated in the sense discussed in Sect.~\ref{Sect::LS_SpatDyn}. This situation leads to a different case in which one must take into account the combined effect of the real doublet and the complex quartet to describe the spatial dynamics.
At some point this combined effect will lead to oscillations in the front profile. 
Eventually there will be a crossover, $q_1=q_2$, after which it is the quartet that determines the asymptotic approach. The onset of oscillations in the spatial profile is not as clear-cut as the transitions described by collisions of spatial eigenvalues described before. The mathematical crossover $q_1=q_2$ indicates the region where the complex quartet becomes as important for the spatial dynamics as the real doublet and therefore fronts have oscillatory tails. However, fronts develop oscillatory tails before the mathematical crossover of eigenvalues.

Both the crossover and the BD transition, result in the front tails going from monotonic to oscillatory decay. However there is a distinctive difference in the front profile resulting from the two transitions, at the BD transition the complex quartet arises with zero imaginary part, thus the wavelength of the oscillation is initially infinite, while in the case of the crossover the complex quartet has a finite imaginary part and thus the wavelength is finite. 

We note that the collision of two complex quartets, which would involve 8 spatial eigenvalues, results in two complex quartets and therefore there is no qualitative change in the spatial dynamics. Thus in what follows the only collisions that we will consider are the ones involving doublets, since they are associated to transitions in the spatial dynamics.

In Part II, when considering the Mexican-hat kernel with positive and negative regions, we will encounter a crossover transition involving six spatial eigenvalues. On the other hand, the other two kernels which are positive definite only have HH or BD transitions that can be explained with just 4 eigenvalues. In the forthcoming sections we identify the codim-2 points that organize the BD, MI and crossover transitions by considering a six order dispersion relation which is the minimal one that can account for six spatial eigenvalues. In terms of $u=-\lambda^2$, this is a cubic $\tilde \Gamma (u)$
\begin{equation}
 \tilde \Gamma (u)= \mu + \alpha u + \beta u^2 - u^3 \,
 \label{Eq::cubic_dispersion_relation}
\end{equation}
In Sect.~\ref{Sect::conclusions} we will relate the coefficients $\alpha$ and $\beta$ to the nonlocal kernel parameters. The sign of the cubic term is taken so that when considering temporal dynamics, $u=k^2$, large wavelength perturbations are damped.

\section{The quadruple zero {\em point}}
\label{Sect::QZ}

As described before, at the BD and HH transitions 
$\tilde \Gamma(u)$ has a DZ at a real value $u_c$ which corresponds to a pair of DZ of $\Gamma_s(\lambda)$ at $\lambda_c=\pm\sqrt{-u_c}$ signaling the collision of two doublets. 
Considering the space of parameters, the manifold defined by a real double zero (RDZ) of $\tilde \Gamma(u)$ (which we will refer as RDZ manifold) is of codim-$1$. The part of the RDZ manifold where $u_c>0$ corresponds to a HH while where $u_c<0$ corresponds to a BD. Since HH and BD are part of the same manifold, they can be seen as the continuation of each other. The conversion from one to the other occurs at the so called quadruple zero (QZ) point \cite{Champneys98,Haragus11} where $u_c=0$. The name comes from the fact that for $u_c=0$ the two DZ of $\Gamma_s(\lambda)$ coincide, thus this point is indeed a quadruple zero of $\Gamma_s(\lambda)$ [though not of $\tilde{\Gamma}(u)$].
In the space of parameters the QZ ``point'' is in fact a codim-$2$ manifold. 
For a detailed description we refer to Ref. \cite{Champneys98} or to section 4.3.5 of Ref. \cite{Haragus11} (where the QZ is referred as reversible $0^{4+}$ bifurcation). Here we will mainly focus on the implications of the unfolding in the existence of LSs in different parameter regions.

The QZ point can be unfold considering four spatial eigenvalues given by the zeros of a dispersion relation quartic in $\lambda$. Here we will make use of the six order dispersion relation (cubic in $u$) (\ref{Eq::cubic_dispersion_relation}) since the QZ will turn out to be part of a broader scenario to be described in the forthcoming sections.
Applying conditions (\ref{Eq::DZ_conditions_u}) to Eq.~(\ref{Eq::cubic_dispersion_relation}) one gets the RDZ manifold which is given by
\begin{align}
 0&=\mu + \alpha u_c + \beta u_c^2 - u_c^3  \label{Eq::cubic_RDZ_condition} \\
 0&= \alpha + 2 \beta u_c - 3u_c^2 .
 \label{Eq::cubic_RDZ_condition_derivate}
\end{align}
From (\ref{Eq::cubic_RDZ_condition_derivate}) one has
\begin{equation}
  u_c=\frac{\beta}{3} \pm \frac{\sqrt{3\alpha+\beta^2}}{3} \,. 
  \label{Eq::cubic_uc}
\end{equation}
Substituting this into (\ref{Eq::cubic_RDZ_condition}) one obtains that in the three dimensional parameter space $(\mu,\alpha,\beta)$ the RDZ manifold is the surface given by
\begin{equation}
 \mu_{\rm RDZ}=-\frac{\beta}{27}(9\alpha+2\beta^2) \mp \frac{2}{27} (3\alpha+\beta^2)^{3/2} \,.
 \label{Eq::cubic_muRDZ}
\end{equation}
The QZ point, $\tilde \Gamma (0)=\tilde \Gamma' (0)=0$ is given by
\begin{equation}
 \mu_{\rm QZ}=0 \, ; \quad \alpha_{\rm QZ} =0 \, ,
 \label{Eq::cubic_QZ}
\end{equation}
Thus in parameter space the QZ manifold is a line with $\mu=\alpha=0$, while $\beta$ is arbitrary.

\begin{figure}[t!]
\includegraphics{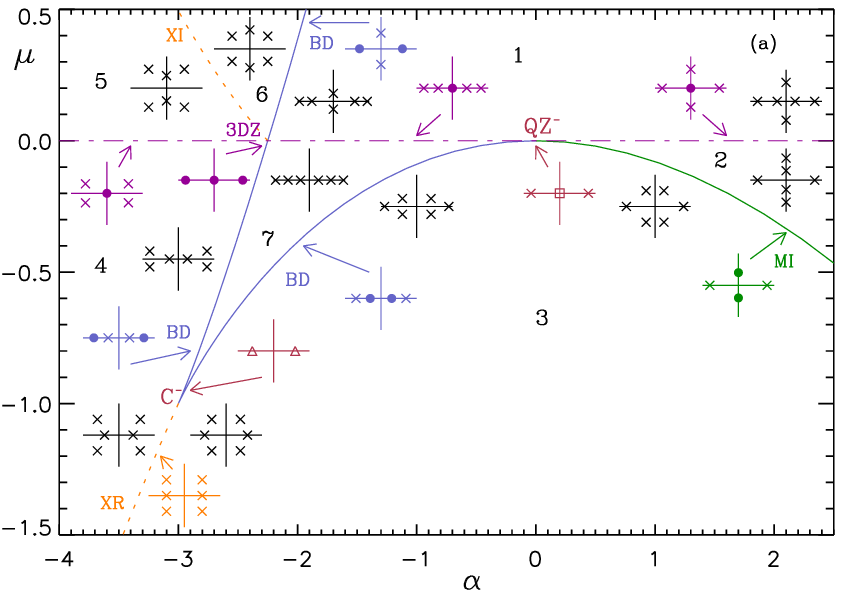}
\includegraphics{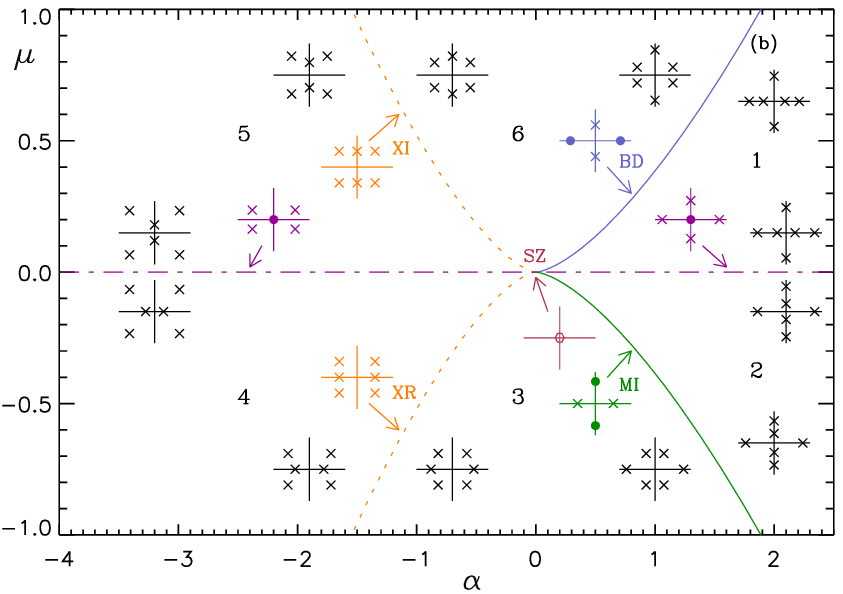}
\includegraphics{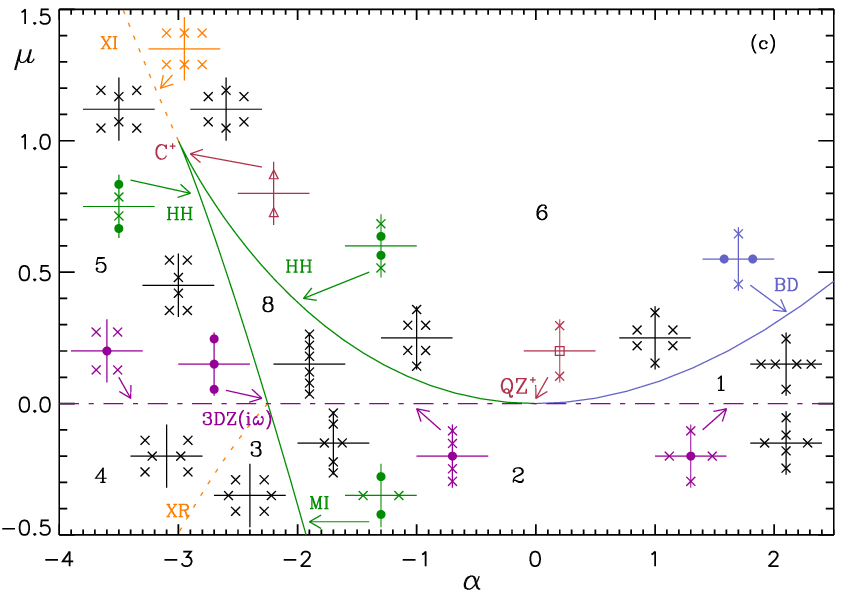}
 \caption{(Color online) Boundaries in the $(\mu,\alpha)$ parameter space at which the spatial eigenvalues of (\ref{Eq::cubic_dispersion_relation}) exhibit different transitions
 for: (a) $\beta=-3$, (b) $\beta=0$ and (c) $\beta=3$. Sketches indicate the position of the zeros of $\Gamma_s(\lambda)$ in the (${\rm Re}(\lambda),{\rm Im}(\lambda)$) plane ($\times$ signal simple zeros, $\CIRCLE$ double zeros, $\triangle$ triple zeros, $\square$ quadruple zeros, and $\hexagon$ sextuple zeros).}
 \label{Fig::SZ}
\end{figure}

Figure \ref{Fig::SZ}(a) shows the different regions in the $(\mu,\alpha)$ parameter space for $\beta=-3$. The insets sketch the location of the spatial eigenvalues in the $({\rm Re}(\lambda),{\rm Im}(\lambda))$ plane. The QZ is located at the origin and for the purposes of this section we just focus on the parameter region close to the QZ. The other parts of the figure will be discussed in the next sections. Also for later convenience we will label this QZ as QZ$^-$ referring to the fact that for $\beta=-3$, $\tilde \Gamma''(0)<0$. The RDZ manifold that unfolds from QZ$^-$ is given by Eq.~(\ref{Eq::cubic_muRDZ}) with the $-$ sign (the other branch will be relevant in Sect.~\ref{Sect::cusp}).
The part of the RDZ manifold that unfolds at the left of QZ$^-$ has a negative value of $u_c$ and therefore corresponds to a BD while the part that unfolds on the right corresponds to a HH. In the region below the HH and BD lines, labeled as $3$, the leading spatial eigenvalues are a complex quartet, therefore fronts connecting HSS will have oscillatory tails leading to the possibility of formation of LSs. When crossing the BD line from region 3 one enters in region 7 where the leading spatial eigenvalues are a real doublet. The fronts connecting two HSS are monotonic and therefore LSs are not formed. Considering the temporal dynamics in regions 3 and 7 the relation $\Gamma (k)$ is negative for all $k$ (and thus the HSS is stable).

\begin{figure}
 \includegraphics{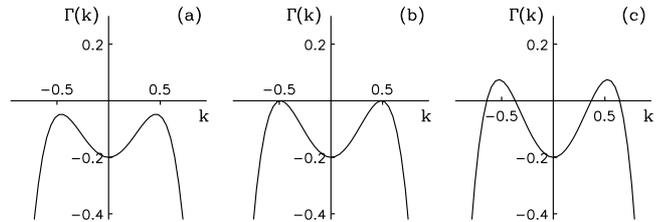}
 \caption{Dispersion relation $\Gamma(k)$ for $\beta=-3$, $\mu=-0.2$ and (a) $a=1.4$ (region 3), (b) $a=1.6109$  (on the MI line), and (c) $a=1.9$ (region 2).}
 \label{Fig::MI32}
\end{figure}

When crossing the HH one enters in region 2 where the leading spatial eigenvalues are two imaginary doublets $\lambda_1=\pm i k_1$ and $\lambda_2=\pm i k_2$. As a consequence $\Gamma(k)>0$, for  $k_1<\abs{k}<k_2$, so that the HSS is unstable to perturbations with wavenumber within that range, as shown in Fig.~\ref{Fig::MI32}. Thus, the HH unfolding from QZ$^{-}$ corresponds indeed to a MI of the HSS. In region 2, since the HSS is unstable, no stable LSs can be formed.

\begin{figure}
 \includegraphics{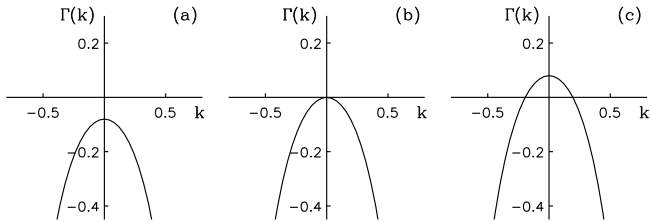}
 \caption{Dispersion relation $\Gamma(k)$ for $\beta=-3$, $a=-2$ and (a) $\mu=-0.08$ (region 7), (b) $a=0$  (on the Hamiltonian-pitchfork line), and (c) $a=0.08$ (region 1).}
 \label{Fig::RTB71}
\end{figure}

The last of the regions surrounding QZ$^-$ is region 1 is separated from 2 and 7 by transitions 
different from the ones considered in Sect.~\ref{Sect::transitions_LS} since there we only discussed the transitions which bring the system into the region where LSs can exist (region 3 in the figure). Both transitions involve the collision of the two components of a doublet located on the real axis leading to the formation of a doublet on the imaginary axis. This is a DZ of $\Gamma(k)$ at $k=0$, which, as discussed in Sect.~\ref{Sect::linstab}, is associated to the effect of homogeneous perturbations to the HSS. For the cubic kernel this DZ takes place at $\mu=0$. In region 1, the leading spatial eigenvalues are an imaginary doublet $\lambda_2= \pm i k_2$. From the point of view of the temporal dynamics, $\Gamma(k)>0$ for $-k_2<k<k_2$ [c.f. Fig.~~\ref{Fig::RTB71} (c)], thus the HSS is unstable to perturbations with wavenumber smaller than $k_2$.

\begin{figure}
 \includegraphics{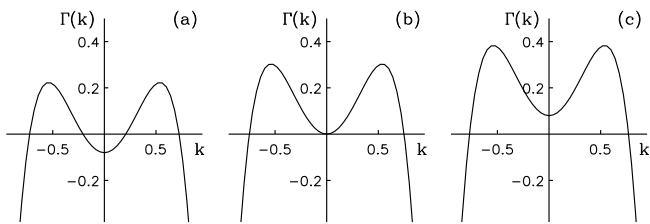}
 \caption{Dispersion relation $\Gamma(k)$ for $\beta=-3$, $a=2$ and (a) $\mu=-0.08$ (region 2), (b) $a=0$  (on the Hamiltonian-pitchfork-Hopf line), and (c) $a=0.08$ (region 1).}
 \label{Fig::RTH21}
\end{figure}

The difference between the transitions from 7 to 1 and from 2 to 1 is given by the location of the spatial eigenvalues not involved in the collision. From 7 to 1 the accompanying eigenvalues are on the real axis and the bifurcation is known as reversible Takens-Bogdanov or Hamiltonian pitchfork or $0^{2+}$ \cite{Champneys98,Haragus11}. The HSS goes from being stable with monotonic fronts to being unstable to small wavelength perturbations. Fig.~\ref{Fig::RTB71} shows the change in $\Gamma(k)$ when crossing this bifurcation.
The transition from 2 to 1 is a reversible Takens-Bogdanov-Hopf or Hamiltonian pitchfork-Hopf \cite{Champneys98} or $0^{2+}(i\omega)$ \cite{Haragus11}, characterized by a pair of imaginary spatial accompanying eigenvalues $\lambda_2=\pm i k_2$. From $2$ to $1$ the components of the imaginary doublet closer to the origin, $\lambda_1=\pm i k_1$, collide leading to a real doublet. Thus, while in region 2 the HSS was unstable only to perturbations with wavenumber $k_1<\abs{k}<k_2$, now it becomes unstable to a wider range $\abs{k}<k_2$, which includes homogeneous perturbations (see Fig.~\ref{Fig::RTH21}).

\section{The cusp {\em point}}
\label{Sect::cusp}

Consider the collision of two conjugate complex zeros of $\tilde \Gamma(u)$ on the real axis for $u$ leading to two real zeros, as in the HH and BD transitions, but now the location where the collision takes place $u_{\rm cusp}$ coincides with a simple real zero of $\tilde \Gamma (u)$. This is a codim-2 point at which
$\tilde \Gamma (u)$ has, by definition, a triple zero: $\tilde\Gamma(u_{\rm cusp})=\tilde\Gamma'(u_{\rm cusp})=\tilde\Gamma''(u_{\rm cusp})=0$. 
After the collision one of the complex zeros of $\tilde \Gamma(u)$ pairs with the real zero to form a RDZ while the other complex zero becomes a simple real zero. There are two ways in which this pairing can be done, therefore in parameter space the outcome are two RDZ manifolds emerging one tangent to the other, forming a cusp. On one of the emerging RDZ manifolds $\tilde\Gamma''(u)<0$,  which corresponds to a local maximum in $\tilde\Gamma(u)$, while, conversely on the other $\tilde\Gamma''(u)>0$, corresponding to a local minimum. 

If $u_{\rm cusp}>0$ the two RDZ manifolds unfolding from the cusp correspond to a HH bifurcation and we label the cusp as C$^+$, while if $u_{\rm cusp}<0$ the cusp unfolds two BD manifolds and we label it as C$^-$.

In terms of $\Gamma_s(\lambda)$ the cusp corresponds to two triple zeros located at $\lambda_{\rm cusp}=\pm\sqrt{-u_{\rm cusp}}$. 
Considering the location of the spatial eigenvalues in the $({\rm Re}(\lambda),{\rm Im}(\lambda))$ plane, at a C$^-$ the components of a complex quartet collide on the real axis on the same location where there is a real doublet. Conversely, at a C$^+$ the components of a complex quartet collide on the imaginary axis on the same location where there is a imaginary doublet.

Cusp points in reversible systems have been thoroughly studied in the context of traveling waves in Fermi-Pasta-Ulam lattices \cite{Iooss00}. These cusps, also known as reversible $0^{3+}$ bifurcations \cite{Haragus11}, arise in odd-dimensional reversible systems. They are of codim-1 and involve three spatial eigenvalues. At the cusp a pair of complex conjugated spatial eigenvalues collide with a real spatial eigenvalue. For the even-dimensional reversible systems considered in this work due to the symmetry in the location of the spatial eigenvalues, this event can not take place. Instead we have codim-2 cusps of BDs or HHs which involve two triple zeros of the dispersion relation and six spatial eigenvalues. 

We now focus on the consequences for the existence of LSs that follow from the unfolding of the cusp of BDs or HHs. Setting $\tilde \Gamma''(u_{\rm cusp})=0$ in the dispersion relation (\ref{Eq::cubic_dispersion_relation}) one has,
\begin{equation}
 u_{\rm cusp}=\frac{\beta}{3} \, .
\end{equation}
Therefore for $\beta<0$ the cusp is a C$^-$ one, while for $\beta>0$ is C$^+$. Its location can be obtained setting $\tilde \Gamma'(u_{\rm cusp})=\tilde \Gamma(u_{\rm cusp})=0$, which leads to
\begin{equation}
 \alpha_{\rm cusp}=-\frac{\beta^2}{3}, \quad \mu_{\rm cusp}=-\frac{\beta^3}{27} \, .
\end{equation}

Figure \ref{Fig::SZ}(a) shows the C$^-$ cusp unfolding two BD lines which correspond to the two possible signs in Eq.~(\ref{Eq::cubic_muRDZ}). The region between the two BDs corresponds to region 7 which has three real doublets and, as described in Sect.~\ref{Sect::QZ}, fronts connecting HSS are monotonic. Beyond the cusp, one encounters a large area in parameter space in which there is a real doublet $\lambda_1=\pm q_1$ plus a complex quartet $\lambda_2=\pm q_2 \pm i k_2$ (regions 3 and 4). In region 3, $q_1>q_2$, thus the complex quartet leads the spatial dynamics, while in region 4, $q_1<q_2$, so that the spatial dynamics is lead by a real doublet.
The separation between region 3 and 4 is given by the crossover manifold discussed in Sect.~\ref{Sect::transitions_LS}, whose location can be determined by writing the dispersion relation $\Gamma_s(\lambda)$ as function of its zeros,
\begin{align}
\Gamma_s (\lambda) = & (\lambda+q_1)(\lambda-q_1) (\lambda+q_2+ik_2) (\lambda+q_2-ik_2) \nonumber \\ 
& \times (\lambda-q_2+ik_2)(\lambda- q_2-ik_2) \, .
\end{align}
Setting $q_1=q_2$ and considering $u=-\lambda^2$ one gets,
\begin{align}
\tilde \Gamma(u)=&-u^3+ (2 k_2^2 -3 q_2^2) u^2 - (k_2^4+3 q_2^4) u \nonumber \\
&- q_2^2 \left(k_2^2+ q_2^2 \right)^2 .
\end{align}
Comparing with (\ref{Eq::cubic_dispersion_relation}) one has,
\begin{equation}
 \beta=2 k_2^2-3q_2^2, \; 
 \alpha=-k_2^4-3q_2^4,   \;
 \mu=-q_2^2 \left(k_2^2+ q_2^2 \right)^2 .
\end{equation}
Eliminating $q_2$ and $k_2$ one obtains,
\begin{equation}
 \mu_{\rm XR}=-\frac{\beta(35\alpha+2\beta^2)}{1029}+\frac{(350 \alpha + 74 \beta^2)}{1029 \sqrt{3}}\sqrt{-7\alpha-\beta^2} .
 \label{Eq::cubic_muXR}
\end{equation}
Notice that this expression signals a crossover only if $\alpha$ and $\beta$ are such that $q_s^2>0$ and $k_2^2>0$. Fig.~\ref{Fig::SZ}(a) shows $\mu_{\rm XR}$ as it unfolds from C$^-$.

As discussed in Sect.~\ref{Sect::transitions_LS}, for parameter values in the part region 4 close to the crossover one may expect to encounter oscillatory tails which can allow for the formation of LS. Within region 4 going away from the crossover the complex quartet will move away from the imaginary axis and eventually its contribution to the spatial dynamics will be irrelevant and fonts will be monotonic.

\begin{figure}
 \includegraphics{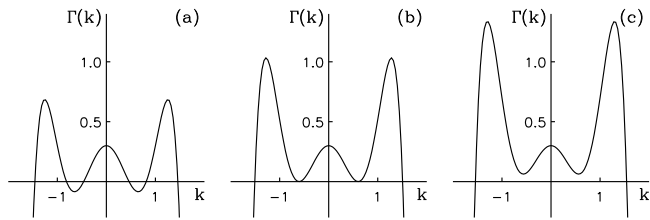}
 \caption{Dispersion relation $\Gamma(k)$ for $\beta=3$, $\mu=0.3$ and (a) $a=-2.0$ (region 8), (b) $a=-1.7836$  (on the HH line), and (c) $a=-1.6$ (region 6).}
 \label{Fig::HH86}
\end{figure}

Figure \ref{Fig::SZ}(c) for $\beta=3$ shows a C$^+$ cusp unfolding two HH lines associated to the two possible signs in Eq.~(\ref{Eq::cubic_muRDZ}). The cusp can be seen as the collision of two HH manifolds after which there is no HH manifold. In the region between the two HHs, labeled as 8, there are three imaginary doublets at $\lambda_1=\pm i k_1$, $\lambda_2=\pm i k_2$ and $\lambda_3=\pm i k_3$. From the temporal point of view the dispersion relation $\Gamma(k)$ is positive, and thus the HSS unstable, for $k$ close to the origin, $\abs{k}<k_1$, and in the range $k_2 < \abs{k}<k_3$ [see Fig.~\ref{Fig::HH86} (a)]. When crossing the HH line to enter in region 6, the two doublets located closer to the origin collide leading to a complex quartet. In fact, the HH corresponds to a minimum of $\Gamma(k)$ crossing zero as shown in Fig.~\ref{Fig::HH86} (thus it does not signal a MI).
As a result in region 6, $\Gamma(k)$ is positive for $\abs{k}<k_3$, thus the HSS is unstable to perturbations with wavenumber in that range.

\begin{figure}
 \includegraphics{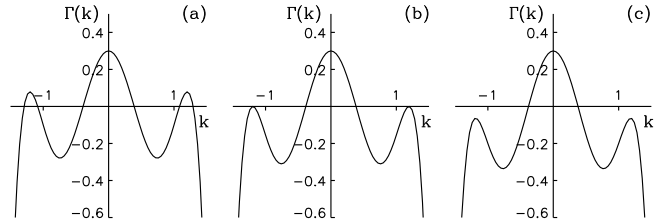}
 \caption{Dispersion relation $\Gamma(k)$ for $\beta=3$, $\mu=0.3$ and (a) $a=-2.4$ (region 8), (b) $a=-2.4549$  (on the HH line), and (c) $a=-2.5$ (region 5).}
 \label{Fig::HH85}
\end{figure}

When crossing the HH line from region 8 to 5 the two external doublets collide, leading to a complex quartet. The HH indeed corresponds to a maximum of $\Gamma(k)$ becoming positive but it is not the global maximum which is located 
at $k=0$ as shown in Fig.~\ref{Fig::HH85} (thus, neither this HH signals an MI).
Therefore in region 5, $\Gamma(k)$ remains positive for (and the HSS unstable to) small wavenumbers $\abs{k}<k_1$.

The C$^+$ generically unfolds a codim-1 manifold signaling the crossover of the imaginary parts of the doublet and the quartet. Proceeding in a similar way as before one gets that this crossover is located at
\begin{equation}
 \mu_{\rm XI}=-\frac{\beta(35\alpha+2\beta^2)}{1029}-\frac{(350 \alpha + 74 \beta^2)}{1029 \sqrt{3}}\sqrt{-7\alpha-\beta^2} .
 \label{Eq::cubic_muXI}
\end{equation}
In Fig. \ref{Fig::SZ}(c) this crossover separates region 5 from 6. From the perspective of determining the regions of existence of stable LSs, this crossover has no effect since at both sides the tails are oscillatory and the HSS itself is modulationally unstable.

\section{The sextuple zero codim-$3$ {\em point}}
\label{Sect::SZ}

If the cusp bifurcation discussed in the previous section takes place at $u_{\rm cusp}=0$, the two triple zeros of $\Gamma_s(\lambda)$ coincide, leading to a 6$^{th}$-order zero, therefore we will refer to this point as {\it sextuple zero} (SZ). In parameter space the SZ is a codim-3 point located at $\tilde \Gamma(0)=\tilde \Gamma'(0)=\tilde \Gamma''(0)=0$. For (\ref{Eq::cubic_dispersion_relation}) the SZ point is located at
\begin{equation}
 \mu_{\rm SZ}=0, \quad \alpha_{\rm SZ}=0, \quad \beta_{\rm SZ}=0,
\end{equation}
as shown in Fig.~\ref{Fig::SZ}(b) for $\beta=0$. In fact the cubic dispersion relation (\ref{Eq::cubic_dispersion_relation}) is the minimal one displaying a SZ point. 

The SZ point can be seen as the collision of a cusp with a QZ (compare Fig.~\ref{Fig::SZ}(a) and (b)). 
At the SZ the C$^+$ manifold becomes a C$^-$ and viceversa, thus C$^+$ and C$^-$ manifolds can be seen as continuation of each other. As a consequence they emerge from the SZ in opposite directions (C$^-$ towards $\beta<0$ and C$^+$ towards $\beta>0$). 

In a similar way, from the SZ two QZ codim-$2$ manifolds unfold in opposite directions. The two QZ differ in the sign of $\tilde \Gamma''(0)$, which is negative for QZ$^-$ (unfolding for $\beta<0$) and positive for QZ$^+$ (unfolding for $\beta>0$). At the same time, the sign of $\tilde \Gamma''(0)$ determines the direction in parameter space in which the RDZ unfolds from the QZ. This is the reason why in Figs.~\ref{Fig::SZ}(a) and (c) the RDZs unfold in opposite directions.

Close to the QZ, $\tilde \Gamma''(0)$ gives the coefficient of the $k^4$ term of the temporal dispersion relation.
For QZ$^-$ the coefficient is negative and therefore this term damps large wavenumber perturbations. On the contrary for QZ$^+$, the quartic term in $k$ amplifies large wavenumber perturbations which can lead to instabilities. Physically these instabilities must be compensated by higher order terms, therefore a QZ$^+$ can only exist in physical systems whose dispersion relation is at least cubic in $u$ (six order in $k$).

\begin{figure}
 \includegraphics{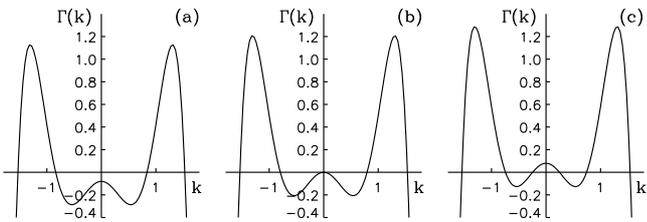}
 \caption{Dispersion relation $\Gamma(k)$ for $\beta=3$, $a=-1.5$ and (a) $\mu=-0.08$ (region 2), (b) $\mu=0$, and (c) $\mu=0.08$ (region 8).}
 \label{Fig::SRTB28}
\end{figure}

Fig.~\ref{Fig::SZ}(c) illustrates the unfolding from QZ$^+$. As for QZ$^-$, in between the HH and the BD there is a doublet and a complex quartet. However here the doublet is located on the imaginary axis and thus this is region 6 in which the HSS is temporally unstable as discussed in Sect.~\ref{Sect::cusp} when describing C$^{+}$. Region 8 as well as the HH line separating it from region 6 were also discussed in Sect.~\ref{Sect::cusp}. Crossing the BD line from region 6 the complex quartet becomes a pair of real doublets leading to region 1, in which HSS are temporally unstable as discussed in Sect.~\ref{Sect::QZ}. Region 2 and the transition from 1 to 2 were also discussed there. Finally, the transition from 2 to 8 involves a DZ of $\Gamma(k)$ at $k=0$, in which a real doublet becomes an imaginary one. In region 8 with three imaginary doublets at $\lambda_1=\pm i k_1$, $\lambda_2=\pm i k_2$ and $\lambda_3=\pm i k_3$, $\Gamma(k)$ is positive for $k$ close to the origin, $\abs{k}<k_1$, and in the range $k_2 
< \abs{k}<k_3$. In region 2, $\Gamma(k)$ is positive only in the range  
$k_2 < \abs{k}<k_3$. As a consequence, the transition from 2 to 8 is a maxima of $\Gamma(k)$ crossing zero but it is not a true homogeneous instability since the HSS was already unstable to finite wavelength perturbations (see Fig.~\ref{Fig::SRTB28}). Looking only at the 4 spatial eigenvalues closer to the origin, this is a Hamiltonian-pitchfork-Hopf bifurcation, however here we have an additional imaginary doublet. Also here at the bifurcation  $\Gamma(k)$ has a local maxima crossing the origin while for the  standard Hamiltonian-pitchfork-Hopf is a minima (compare Fig.~\ref{Fig::SRTB28} with Fig.~\ref{Fig::RTH21}). 

Region 3, in which the spatial dynamics is lead by a complex quartet, also exists for $\beta>0$, as shown in Fig.~\ref{Fig::SZ}(c). Interestingly enough, in this case region 3 does not arise from the unfolding of the QZ. It originates from another codim-2 point, which itself unfolds from the SZ. In terms of the dispersion relation $\Gamma_s(\lambda)$ it corresponds to three DZ, one at the origin and the other two at $\lambda_c=\pm i k_c$.
This implies the collision of two imaginary doublets (a HH) and the collision of the two components of a doublet at the origin (a Hamiltonian-pitchfork-Hopf) taking place simultaneously. Since the collision of the doublets is on the imaginary axis we will refer to this transition as 3DZ$(i\omega)$ (see Fig.~\ref{Fig::SZ}(c)).
The Hamiltonian-pitchfork-Hopf and the HH occur at different places in phase space, thus this is a nonlocal transition.  In terms of $\tilde\Gamma(u)$ it corresponds to the coincidence of a simple zero at the origin $\tilde \Gamma(0)=0$ and a RDZ at finite distance on the positive semi-axis $u_c=k_c^2>0$, $\tilde \Gamma(u_c)=\tilde \Gamma'(u_c)=0$, with non zero $u_c$ (if $u_c=0$ then one has a QZ point rather than a 3DZ$(i\omega)$). 
Since $\tilde \Gamma(0)=0$ implies $\mu=0$, the location of the 3DZ$(i\omega)$ can be obtained setting $\mu_{\rm RDZ}=0$ in Eq. (\ref{Eq::cubic_muRDZ}) and looking for solutions associated to a non zero $u_c$. One obtains
\begin{equation}
 \mu_{{\rm 3DZ}(i\omega)}=0\, , \quad \alpha_{{\rm 3DZ}(i\omega)} = -\frac{\beta^2}{4} \, .
 \label{Eq::cubic_3DZ}
\end{equation}

\begin{figure}
 \includegraphics{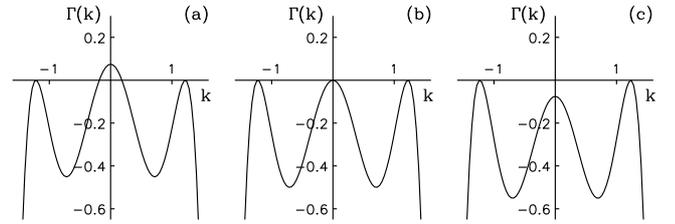}
 \caption{Dispersion relation $\Gamma(k)$ evaluated on top of the HH manifold close to the 3DZ$(i\omega)$ point for $\beta=3$. (a) on the HH line above the 3DZ$(i\omega)$, $a=-2.3$ and $\mu=0.07458$; (b) at the 3DZ$(i\omega)$, $a=-9/4$ and $\mu=0$; and (c) on the MI line below the 3DZ$(i\omega)$ $a=-2.2$ and $\mu=-0.07541$.}
 \label{Fig::HHMI}
\end{figure}

As discussed in Sect.~\ref{Sect::cusp}, the HH unfolding at the left of C$^{+}$ corresponds to two local maxima of $\Gamma(k)$ at $\pm k_c$ crossing zero, but is not a MI because they are not the global maximum, which is located at the origin. However, at the 3DZ$(i\omega)$ point the maximum at the origin crosses zero and after that $\Gamma(0)$ becomes negative (see Fig.~\ref{Fig::HHMI}). As a consequence at the 3DZ$(i\omega)$ the HH acquires a MI character.
When crossing this MI line from region 2, the two imaginary doublets become a complex quartet, which leads the spatial dynamics entering region 3. Moving away from the MI line, the complex quartet goes away from the imaginary axis and eventually the real doublet leads the dynamics entering in region 4. The crossover manifold is given by Eq. (\ref{Eq::cubic_muXR}) but it does not unfold from the cusp. Instead it unfolds from the 3DZ$(i\omega)$ secant to the MI line, as shown in Fig.~\ref{Fig::SZ}(c).

\begin{figure}
 \includegraphics{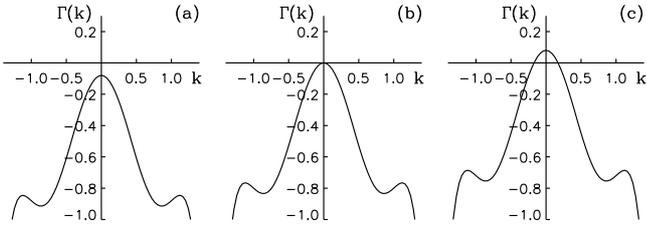}
 \caption{Dispersion relation $\Gamma(k)$ for $\beta=3$, $a=-2.8$ and (a) $\mu=-0.08$ (region 4), (b) $\mu=0$ (on the Hamiltonian-pitchfork line), and (c) $\mu=0.08$ (region 5).}
 \label{Fig::RTB45}
\end{figure}

The 3DZ$(i\omega)$ has also an effect on the manifold of instabilities to homogeneous perturbations located at $\mu=0$. Between QZ$^+$ and 3DZ$(i\omega)$ the collision of the two components of a doublet at the origin is accompanied by two imaginary doublets. Resulting from the HH bifurcation acting on the two imaginary doublets, on the left the 3DZ$(i\omega)$ the accompanying spatial eigenvalues are a complex quartet. This is the case for the transition between regions 4 and 5 illustrated in Fig.~\ref{Fig::RTB45}, which from the temporal point of view is equivalent to a Hamiltonian-pitchfork bifurcation shown Fig.~\ref{Fig::RTB71}. 

A 3DZ point exists also for $\beta<0$, where the BD intersects with the Hamiltonian pitchfork (see Fig.~\ref{Fig::SZ}(a)), the difference being that now the two double zeros at finite distance take place on the real axis of the $({\rm Re}(\lambda),{\rm Im}(\lambda))$ plane. The 3DZ location is also given by the conditions (\ref{Eq::cubic_3DZ}), which are independent of the sign of $\beta$. This point changes the character of the BD line unfolding at the left of C$^-$.
Between C$^-$ and 3DZ the BD is accompanied by a real doublet located closer to the origin and thus leading the spatial dynamics. At the 3DZ this real doublet becomes imaginary and thus above the 3DZ the BD is accompanied by an imaginary doublet. As for the instabilities to homogeneous perturbations at $\mu=0$, between QZ$^-$ and 3DZ the collision of the two components of a doublet at the origin is accompanied by two real doublets. After 3DZ the accompanying eigenvalues are a complex quartet. Nevertheless from the point of view of the temporal dynamics there is not much difference since in both cases the HSS goes from being stable (regions 4 and 7) to being unstable for small wavenumber (regions 5 and 1), thus both transitions correspond to a Hamiltonian-pitchfork bifurcation. 
Finally  3DZ also unfolds a crossover, however in this case the crossover divides regions 5 and 6 which differ in the location of the imaginary part and therefore is irrelevant for the existence of stable LSs as discussed in Sect.~\ref{Sect::cusp}.

In the notation of Ref.~\cite{Haragus11} the SZ point would be referred as reversible $0^{6+}$ bifurcation, and has not been characterized in the literature to the best of our knowledge. Nevertheless this point plays a major role in the overall organization of the spatial dynamics. In the examples that we consider in Part II we will encounter a SZ when considering the Mexican-hat kernel that is not monotonic, exhibiting both attraction and repulsion. The above description just sketches the SZ features that are more relevant for this work. A full description of the point would require a deeper mathematical analysis.

\section{Discussion of the overall scenario and nonlocal kernel effects}
\label{Sect::conclusions}

Spatial dynamics allows to determine the parameter regions in which fronts emerging from a HSS have oscillatory tails and thus where LSs can exist. The presence of oscillatory tails is associated to the fact that the spatial dynamics is lead by a quartet of complex eigenvalues (region 3) or by the combination of a real doublet and a complex quartet (part of region 4).

Moving in parameter space there are three transitions that bring the system into region 3. Two of them correspond to the collision in the $({\rm Re}(\lambda),{\rm Im}(\lambda))$ plane of two doublets leading to a complex quartet: The Hamiltonian-Hopf bifurcation associated to collisions on the imaginary axis, related to the modulational instability of the homogeneous solution, and the Belyakov-Devaney transition corresponding to collisions on the real axis, so that fronts with monotonic tails become oscillatory, initially with infinite wavelength. The third transition arises from a crossover in which starting from a parameter region where the spatial dynamics is lead by a real doublet when changing a parameter a complex quartet moves closer to the imaginary axis bypassing the location of real doublet. As a consequence monotonically decaying fronts acquire oscillations with a finite wavelength. 
The crossover is not a clear-cut transition as in fact oscillatory tails are present prior to the crossover for parameter values where spatial dynamics is lead by a real doublet but with a complex quartet located a similar distance from the imaginary axis (part of region 4).
 
These three transitions unfold from three codim-2 points: the QZ in which the dispersion relation has a quadruple zero, the cusp where two Belyakov-Devaney or two Hamiltonian-Hopf manifolds start (or end), and the 3DZ$(i\omega)$, characterized by three double zeros of the dispersion relation taking place simultaneously. These three codim-2 transitions unfold from the SZ codim-3 local bifurcation point characterized by being a 6$^{th}$ zero of the dispersion relation.

As a consequence of the phase space organization, for $\beta<0$ region 3 unfolds from the QZ, located at $\mu=\alpha=0$, and has a parabolic shape. It is limited, on one side by the MI, and, on the other by the BD for small $\mu$ and by the crossover for $\mu$ beyond the cusp, $\mu>\mu_{\rm cusp}=-\beta^3/27$. When $\beta$ approaches zero region 3 narrows. For $\beta=0$ the QZ collides with the cusp becoming a SZ, still located at $\mu=\alpha=0$ and region 3 is limited by the MI and the crossover. For $\beta>0$ region 3 unfolds from the 3DZ$(i\omega)$ located at $\mu=0$, $\alpha=-\beta^2/4$ and has a sharp-pointed shape limited on one side by the MI, and, on the other side by the crossover. Finally, the part of region 4 where fronts have oscillatory tails is located close to the crossover. 

Nonlocal kernels can bring the system into the parameter regions where fronts have oscillatory tails. To illustrate this consider a system whose dispersion relation without nonlocal coupling is linear in $u$ (quadratic in $\lambda$), 
\begin{equation}
  \tilde \Gamma_G(u)=\mu-a u \,.
\end{equation}
This system has only two eigenvalues and therefore eventual fronts connecting HSS must be monotonic. No LSs can be formed. Now consider a nonlocal kernel whose Fourier transform has no singularities in the complex plane. Since the spatial dynamics is determined by the eigenvalues with smaller real part we can consider a Taylor expansion of the nonlocal kernel around $u=0$ as described in subsection \ref{Sect::MomentExpansion}. For positive definite kernels and $s<0$ the series can, in principle, be truncated at the fourth moment. The spatial dynamics has 4 spatial eigenvalues allowing for BD and MI transitions to occur unfolding from a QZ$^-$ point. 
Thus nonlocal positive definite kernels can lead to oscillatory tails in the front profile for $s<0$. From a physical point of view oscillatory tails arise resulting from the interplay between attractive local interaction and repulsive nonlocal one. We will encounter this situation in Part II when considering the Gaussian kernel. 

Kernels that have attractive and repulsive regions in real space will typically have moments with different signs and show a richer scenario. Consider for instance a kernel of the form (\ref{Eq::scaled_kernel}) in which the moments scale with the width $\sigma$ and that can be expanded up to order $M_6$:
\begin{equation}
\tilde{\hat{K}}_{\sigma}(u) \approx M_0 - \frac{{\cal M}_2 \sigma^2}{2} u + \frac{{\cal M}_4 \sigma^4}{4!} u^2 - \frac{{\cal M}_6 \sigma^6}{6!} u^3 .
\label{Eq::Kernel_cubic}
\end{equation}
The overall dispersion relation (\ref{Eq:Gamma_tilde}) is then given by,
\begin{equation}
 \tilde \Gamma(u)= \mu- \left(a + \frac{s {\cal M}_2 \sigma^2}{2}\right) u + \frac{s {\cal M}_4 \sigma^4}{4!} u^2 - \frac{s{\cal M}_6 \sigma^6}{6!} u^3 .
\end{equation}

Assuming $s{\cal M}_6 \sigma^6 >0$ (which ensures stability to large wavelength perturbations) and defining, 
\begin{equation}
 v=\left( \frac{s{\cal M}_6}{6!}\right)^{1/3}\sigma^2 u \, ,
 \label{Eq::change_variables_v}
\end{equation}
the overall dispersion relation can be written in the same form as (\ref{Eq::cubic_dispersion_relation}), 
\begin{equation}
 \tilde \Gamma(v)=\mu+ \alpha v + \beta v^2 - v ^3  \, ,
 \label{Eq::Gamma_v}
\end{equation}
where,
\begin{align}
 \alpha &= - \left(\frac{90}{s{\cal M}_6}\right)^{1/3} \left( \frac{2a}{\sigma^2} + s{\cal M}_2 \right) 
 \label{Eq::coef_alpha}\\
 \beta  &= 5 \left(\frac{3s}{2{\cal M}_6^2}\right)^{1/3} {\cal M}_4 
 \label{Eq::coef_beta}\, .
\end{align}
For $\mu<0$, playing with the kernel shape (${\cal M}_2$, ${\cal M}_4$ and ${\cal M}_6$), the width $\sigma$ or the strength $s$ allows to change the values of the coefficients $\alpha$ and $\beta$ in order to bring the system into region 3 or in the part of region 4 close to the crossover. In particular kernels which can lead to a negative value for $\beta$ will be more suitable to induce LSs since, as shown in Fig.~\ref{Fig::SZ}, region 3 is larger. The sign of $\beta$ is that of $s{\cal M}_4$. The sign of ${\cal M}_2$ and ${\cal M}_4$ can be changed by varying the weight of the attractive and repulsive parts of the kernel. Particularly interesting is the case in which the second and fourth moment have the opposite sign than the six order one. Then $s{\cal M}_4<0$ and 
Choosing the kernel parameters so that $s \sigma^2 {\cal M}_2=-2a$ one has a negative $\beta$ and $\alpha=0$ where the system is in region 3 for any negative $\mu$. The balance does not need to be perfect, since region 3 is quite large and even if $\alpha \neq 0$ the system can be brought there provided $\mu$ is not too close to zero. 

Nonlocal kernels can also avoid the formation of LSs in systems in which they are present. To illustrate this consider a system whose dispersion relation without nonlocal coupling is quadratic in $u$ (quartic in $\lambda$),
\begin{equation}
  \tilde \Gamma_G(u)= \mu + a u - b u^2\, ,
\end{equation}
such as for example the Swift-Hohenberg equation. In this case the local dynamics has a QZ point at $\mu=0$, $a=0$ which unfolds a RDZ manifold located at $\mu_{\rm RDZ}=a^2/4b$. For $a<0$ this manifold corresponds to a BD while for $a>0$ to a MI. In the region below the RDZ manifold the spatial dynamics is dominated by a complex quartet and thus LSs exist. We now consider a nonlocal interaction term of the form (\ref{Eq::Kernel_cubic}). By applying the change of variables (\ref{Eq::change_variables_v}) the overall dispersion relation can be written as in (\ref{Eq::Gamma_v}) with
\begin{align}
 \alpha &=  \left(\frac{90}{s{\cal M}_6}\right)^{1/3} \left(  \frac{2a}{\sigma^2} - s{\cal M}_2 \right) \, , \nonumber \\
 \beta  &= 5\left(\frac{3}{2 s^2 {\cal M}_6^2}\right)^{1/3} \left( s  {\cal M}_4 -\frac{4! b}{\sigma^4}\right) \nonumber \, .
\end{align}
For $\mu<0$, it is possible, for example, to bring the system into region 7 where fronts are monotonic by adjusting the kernel shape, its width or its strength. In particular, the width of the kernel $\sigma$ plays a key role in the balance between the contributions arising from the local and nonlocal terms, respectively, in coefficients $\alpha$ and $\beta$. As a result of this balance $\beta$ may take both signs even with monotonic kernels,
for which all the moments $M_2$ and higher have the same sign.

In Part II we will discuss in detail the effect of three nonlocal kernels widely used in the literature on the existence of oscillatory tails in fronts connecting two equivalent HSSs by applying them to the real Ginzburg-Landau equation whose local dynamics leads to monotonically decaying fronts. The kernels considered 
illustrate different cases discussed here: a Gaussian kernel which has a positive definite spatial profile and for which the moment expansion up to fourth order provides reasonably good results, a mod-exponential kernel which despite being positive definite has a singularity in Fourier space and therefore a moment expansion does not work and a Mexican-hat shaped kernel which has attractive and repulsive regions.


\acknowledgments

This work was supported by the Spanish MINECO and FEDER 
under Grants FISICOS (FIS2007-60327), DeCoDicA (TEC2009-14101),
INTENSE@COSYP (FIS2012-30634) and TRIPHOP (TEC2012-36335), by Comunitat Aut\`onoma de les Illes Balears,
and by the Belgian Science Policy Office under grant No.\
IAP-7-35. LG acknowledges support by the
Research Foundation - Flanders (FWO). We thank Prof.~E.~Knobloch for interesting
discussions.

\bibliography{Nonlocal}

\end{document}